\documentclass{iopart}
\usepackage{iopams}
\expandafter\let\csname equation*\endcsname\relax
\expandafter\let\csname endequation*\endcsname\relax
\usepackage{amsmath,amsfonts,amssymb,amsthm}
\usepackage{amsmath,amsfonts,amssymb,amsthm}
\usepackage{txfonts}
\usepackage{graphicx}
\usepackage{enumerate}
\usepackage{braket}
\makeatletter
\renewcommand{\theequation}{%
\thesection.\arabic{equation}}
\@addtoreset{equation}{section}
\makeatother

\usepackage{etoolbox}

\makeatletter
\newcommand{\mainmatter}{%
  \setcounter{footnote}{0}%
  \patchcmd{\@makefntext}{\fnsymbol}{\arabic}{}{}%
  \patchcmd{\@thefnmark}{\fnsymbol}{\arabic}{}{}%
  \def\@makefnmark{\textsuperscript{\arabic{footnote}}}%
}
\makeatother

\begin{document}

\title[]{Multi-soliton solutions of the sine-Gordon equation with elliptic-function background}

\setcounter{footnote}{6}
\author{Daisuke A. Takahashi$^{1,2}$\footnote{The current primary affiliation is 2.}}

\address{
$^1$Research and Education Center for Natural Sciences, Keio University, Hiyoshi 4-1-1, Yokohama, Kanagawa 223-8521, Japan \\ $^2$Department of Physics, Chuo University, 1-13-27 Kasuga, Bunkyo-ku, Tokyo 112-8551, Japan
}
\ead{takahashi@phys.chuo-u.ac.jp}
\begin{abstract}
The multi-soliton solution of the sine-Gordon equation in the presence of elliptic-function background is derived by the inverse scattering method. The key tool in our formulation is the Lax pair written by $4\times4$ matrix differential operators given by Takhtadzhyan and Faddeev in 1974, which enables us to use the conventional form of the integral representation of the Jost solutions and Krichever's theory of commuting differential operators. As a by-product we also provide generalized orthogonality and completeness relations for eigenfunctions associated with indefinite inner product. The multi-soliton solution is expressed by a determinant of theta functions and the shift of the background lattice due to solitons is also determined using addition formula.  One kink and one breather solutions are presented by animated gifs.
\end{abstract}

%\noindent{\it Keywords}:

%\submitto{\jpa}
%\maketitle
\mainmatter
\section{Introduction}
	The sine-Gordon (SG) equation has a wide variety of applications in physics and other natural sciences. For example, it appears as a continuum limit of the Frenkel-Kontorowa model describing the commensurate-incommensurate transition (Ref.~\cite{LambSoliton}, chapter 6 and Ref.~\cite{ChaikinLubensky}, chapter 10), the effective model for the chiral magnet and soliton propagation on it \cite{PhysRevLett.108.107202,PhysRevB.79.134436}, and kinks in the long Josephson junction of superconductors \cite{Fulton1973,PhysRevB.25.5737,doi:10.1143/JPSJ.66.1445}. The model where the partial derivative is changed from hyperbolic to elliptic type has also been solved and applied to vortices and dislocations in spatially two-dimensional systems \cite{Borisov1986,Borisov1988,Borisov1989}. The rogue-wave solutions with unstable background has been recently investigated \cite{Pelinovsky2020}. From the view of the theory of classical integrable systems, the SG equation is one of the earliest equations formulated via the zero-curvature expression, known as the Ablowitz-Kaup-Newell-Segur (AKNS) formalism \cite{PhysRevLett.30.1262,AKNS1974}. The analogical integrable models with higher group symmetries have been also investigated (Ref.~\cite{BudTak77} and references found in Ref.~\cite{FaddeevTakhtajan}, part II, chapter I, \S 8).\\
	\indent In some of the above-mentioned works, the multi-soliton excited states not only for uniform background but also for oscillating non-uniform background attract considerable physical attention. While the multi-soliton solutions of an integrable equation in the presence of the elliptic-function, or more general quasiperiodic Riemann theta function background, can be in principle obtained by taking a special limit of the finite-zone quasiperiodic solutions \cite{BelokolosBobenkoEnolskiiItsMatveev}, the procedure is often complicated. Furthermore, in physical applications, the eigenfunctions of the Lax pair, which is necessary in construction of soliton solutions, often play additional roles in, e.g., calculation of dispersion relations of linearized waves and linear stability analysis. Therefore, constructing the soliton solutions with non-uniform background not from a degenerate case of quasiperiodic solutions but by adding solitons to the stationary background using eigenfunctions of the Lax pair via various conventional methods provides helpful by-products in investigation of physical phenomena. \\
	\indent Motivated by these circumstances, in this paper, we derive the multi-soliton solution of the SG equation in the presence of elliptic-function background. Though this problem itself has been solved in many references including the above-mentioned ones, we believe that the following features (i)-(iii) will bring some new light. (i) We use the Lax formalism based on $4\times 4$ matrix differential operator by Takhtadzhyan and Faddeev \cite{Takhtadzhyan1974}, which is a variant of the Lax pair written by an integral operator \cite{1974JETP...39..228T}. While the famous treatment of the SG equation in soliton theory is the zero-curvature expression using $2\times 2$ matrices which depend on the spectral parameter \cite{PhysRevLett.30.1262,AKNS1974}, the use of this method has the following advantages: (a) the theory of commuting differential operators \cite{Krichever1977} can be applied and (b) the common integral representation of Jost solutions can be used without modification so we do not need trial and error to find suitable form. (ii) The final expression of the multi-soliton solution, which is obtained by formulating the inverse scattering method (ISM) and solving the Gelfand-Levitan-Marchenko (GLM) equation, is compactly summarized as a determinant of theta functions, and its asymptotic form is also derived using the addition formulas in Ref.~\cite{Kharchev201519}. (iii) The orthogonal and completeness relations of eigenfunctions arising from the indefinite inner product is discussed in detail, which is necessary in formulation of the ISM and determination of the possible emerging patterns of discrete eigenvalues corresponding to solitons. This treatment is regarded as a generalization of the corresponding finite-dimensional linear algebra \cite{Takahashi2015101} to continuous space and differential operators. \\
	\indent The organization of this paper is as follows. In Sec.~\ref{sec:laxpair}, we write down the Lax pair of the SG equation. In Sec.~\ref{sec:laxsymsigmaortho}, we point out that the Lax operators are ``$\sigma$-self-adjoint'', and introduce the associated indefinite inner product which defines the generalized orthogonal relations to eigenfunctions. In Sec.~\ref{sec:solitonlattice}, we derive simultaneous eigenfunctions of the Lax pair for the stationary soliton lattice solution, and identify the corresponding algebraic curve, based on the theory of commuting differential operators \cite{Krichever1977}. The eigenvalues and eigenfunctions are parametrized by uniformization variable on torus, and their symmetries are also presented. In Sec.~\ref{sec:scatmat}, we define left and right Jost solutions and the scattering matrix for the system where the background potential asymptotically tends to the stationary soliton lattice at spatial infinities.  In Sec.~\ref{sec:glm}, we introduce the integral representation of the Jost solution, and present the GLM equation which determines the kernel function of the integral representation from the scattering data. In Sec.~\ref{sec:reflectionlesspot}, we solve the GLM equation for reflectionless case and determine the multi-soliton solution, which is expressed by determinant of theta functions. The phase shift of the background lattice is also found. In Sec.~\ref{sec:timeevo}, by solving the  time-dependent Lax equation, we determine the time-evolution of the scattering matrix, and using this, we derive the time-dependent multi-soliton solution for the SG equation. In Sec.~\ref{sec:realcondition}, we provide the constraint that the discrete eigenvalues must satisfy in order for the resultant solution to become real and bounded. In Sec.~\ref{sec:anime}, we address the gif animations of the soliton solutions generated by Mathematica. The method of visualization and used numerical values are presented. Sec.~\ref{sec:summary} is devoted to summary and discussion. The appendices discuss several technical details. In \ref{app:FAI}, the integration formula necessary to obtain the eigenfunction is presented. In \ref{app:BAfunction}, we show the detailed derivation of the eigenfunctions given in Sec.~\ref{sec:solitonlattice}. In \ref{app:completeness}, we derive the completeness relation for eigenfunctions of the soliton lattice potential. There, we also discuss several technical important points on, e.g., the zero-norm eigenfunction defined by indefinite inner product and the necessity of the expression by meromorphic integrand. In \ref{app:glm}, we provide the detailed derivation of the GLM equation. In \ref{app:dettheta}, we calculate the determinant of theta functions appearing in the asymptotic form of the soliton solution using addition formulas. 

\section{Lax pair}\label{sec:laxpair}
	We write an $ n\times n $ identity and zero matrix as $ I_n $ and $ O_n $. Let $ \sigma_1,\sigma_2, $ and $ \sigma_3 $ be the Pauli matrices. The Lax pair and the Lax equation for the SG equation is given by \cite{Takhtadzhyan1974}
	\begin{align}
		4\mathrm{i}\frac{\partial \hat{L}}{\partial t}&=[\hat{L},\hat{B}], \label{eq:lax01} \\
		\hat{L}&=-4\mathrm{i}\partial_x\begin{pmatrix} \sigma_3 & \\ & O_2 \end{pmatrix}+\begin{pmatrix} \mathrm{i}\sigma_1w & \mathrm{e}^{-\mathrm{i}\phi\sigma_2/2} \\ \mathrm{e}^{-\mathrm{i}\phi\sigma_2/2} & O_2 \end{pmatrix},\quad w= \phi_x+\phi_t, \label{eq:lax02} \\
		\hat{B}&=4\mathrm{i}\partial_x\begin{pmatrix} I_2 & \\ & -I_2 \end{pmatrix}-2\begin{pmatrix} & \sigma_3\mathrm{e}^{-\mathrm{i}\phi\sigma_2/2} \\ \mathrm{e}^{-\mathrm{i}\phi\sigma_2/2}\sigma_3 & \end{pmatrix}. \label{eq:lax03}
	\end{align}
	Equation (\ref{eq:lax01}) then reduces to the SG equation
	\begin{align}
		\phi_{tt}-\phi_{xx}+\sin\phi=0.
	\end{align}
	Here we slightly changed prefactors of operators and choice of the SU(2) basis in the way $ (\sigma_1,\sigma_2,\sigma_3) \to (\sigma_1,\sigma_3,-\sigma_2) $ from the original work Ref.~\cite{Takhtadzhyan1974}. This makes no essential difference in formulation.
\section{Symmetry of the Lax operator and orthogonality of eigenfunctions}\label{sec:laxsymsigmaortho}
	Henceforth we simply write $ \sigma:=\sigma_3\oplus\sigma_3=I_2\otimes \sigma_3 $. Then, $ \hat{L} $ and $ \hat{B} $ are both ``$ \sigma $-self-adjoint''\cite{Takahashi2015101}, i.e., 
	\begin{align}
		\sigma \hat{L}^\dagger \sigma=\hat{L},\quad \sigma \hat{B}^\dagger \sigma=\hat{B}. \label{eq:sigmahermitian}
	\end{align}
	Therefore the orthogonal and completeness relations of eigenfunctions of these operators are defined through the indefinite inner product, called ``$\sigma$-inner product'' in \cite{Takahashi2015101} : $ (f_1,f_2)_\sigma:=\int dx f_1^\dagger \sigma f_2 $. The correct identification of the completeness relation is important in derivation of the GLM equation (\ref{app:completeness} and \ref{app:glm}). \\
	\indent Following the established procedure of the ISM, we first consider the eigenvalue problem of $\hat{L}$. Since the highest-order coefficient matrix of $ \hat{L} $ is given by $ \sigma_3\oplus O_2 $, which obviously has rank 2, there exist two linearly independent eigenfunctions for a given eigenvalue $\lambda$. They can be written in the form $ f=\left(\begin{smallmatrix}g\\ \lambda^{-1}\mathrm{e}^{-\mathrm{i}\phi\sigma_2/2}g\end{smallmatrix}\right) $ with $ g $ a two-component column vector. If we rewrite the equation with respect to $g$ using the light-cone coordinate $ x'=\frac{x+t}{2},\ t'=\frac{x-t}{2} $, the famous $ 2\times 2 $ AKNS form \cite{PhysRevLett.30.1262} is reproduced. Let $ f_1=\left(\begin{smallmatrix}g_1\\ \frac{1}{\lambda}\mathrm{e}^{-\mathrm{i}\phi\sigma_2/2}g_1\end{smallmatrix}\right) $ and $ f_2=\left(\begin{smallmatrix}g_2\\ \frac{1}{\lambda}\mathrm{e}^{-\mathrm{i}\phi\sigma_2/2}g_2\end{smallmatrix}\right) $ be eigenfunctions of $ \hat{L} $ with eigenvalues $ \lambda_1 $ and $ \lambda_2 $. Then, we can show 
	\begin{align}
		4\mathrm{i}\left( f_1^\dagger (I_2\oplus O_2)f_2 \right)_x=(\lambda_1^*-\lambda_2)f_1^\dagger\sigma f_2. \label{eq:constJ}
	\end{align}
	Therefore, if $ \lambda_1^*\ne \lambda_2 $, then we have $ \int dx f_1^\dagger \sigma f_2=0 $, which is the above-mentioned orthogonality. On the other hand, when $ \lambda_1^*=\lambda_2 $, we obtain $ f_1^\dagger (I_2\oplus O_2)f_2=g_1^\dagger g_2=(x\text{-independent}) $. We can also show that if $ \lambda_1=\lambda_2 $, $ \det(g_1,g_2) $ does not depend on $ x $, which can be more easily shown using the AKNS form. \\
	\indent In addition to Eq.~(\ref{eq:sigmahermitian}), $\hat{L}$ further has the following symmetry:
	\begin{align}
		(I_2\otimes \sigma_2)\hat{L}(I_2\otimes \sigma_2)=\hat{L}^*,\quad (\sigma_3\otimes\sigma_2)\hat{L}(\sigma_3\otimes\sigma_2)=-\hat{L},
	\end{align}
	which immediately means
	\begin{alignat}{4}
		&& \hat{L}f&=\lambda f && \quad\leftrightarrow\quad \hat{L}[(I_2\otimes\sigma_2) f^*]&&=\lambda^*(I_2\otimes\sigma_2) f^* \notag \\
		\leftrightarrow \quad && \hat{L}[(\sigma_3\otimes\sigma_2) f]&=-\lambda(\sigma_3\otimes\sigma_2) f  && \quad\leftrightarrow\quad \hat{L}[(\sigma_3\otimes I_2)f^*]&&=-\lambda^*(\sigma_3\otimes I_2)f^*. \label{eq:invol}
	\end{alignat}
	Thus the eigenvalues  $ \lambda,\lambda^*,-\lambda,-\lambda^* $ always appear simultaneously.\\ 
	\indent $ \hat{L} $ also has a little unfamiliar symmetry:
	\begin{align}
		&\hat{L}f=\lambda f \quad\leftrightarrow \quad \tilde{L}[(\sigma_2\otimes\sigma_3)f]=-\lambda^{-1}(\sigma_2\otimes\sigma_3)f, \label{eq:Linv1} \\
		&\tilde{L}:= \text{an operator such that $ w $ in $ \hat{L} $ is replaced by  $ \tilde{w}=2\phi_x-w $.}
	\end{align}
	We thus find the identity
	\begin{align}
		\hat{L}^{-1}=-(\sigma_2\otimes\sigma_3)\tilde{L}(\sigma_2\otimes\sigma_3). \label{eq:Linv3}
	\end{align}
	It is unusual that the inverse of a matrix differential operator can be written down explicitly, and furthermore, its expression does not include an integral operator. Such situation seems to occur when the highest-order coefficient matrix is not full-rank, which is $ -4\mathrm{i}\sigma_3 \oplus O_2 $ for the present $\hat{L}$. \\
	\indent In particular, for the stationary solution of the SG equation $ \phi_t=0 $, we find $w=\tilde{w}=\phi_x$ and hence $\hat{L}=\tilde{L}$. In this case $ \hat{L}^{-1}=-(\sigma_2\otimes\sigma_3)\hat{L}(\sigma_2\otimes\sigma_3) $ holds and the eigenvalues $\lambda$ and $-\lambda^{-1}$ appear in pairs. We, however, again emphasize that this relation does not hold for general $ w \ne \phi_x $, and therefore when we consider the scattering matrix $S$ in the ISM in Sec.~\ref{sec:scatmat}, we must not impose this symmetry to $\hat{L}$.
\section{Soliton lattice solution, Riemann surface, and eigenfunctions}\label{sec:solitonlattice}
	Let us consider the eigenfunction for the stationary solution, which we henceforth write $\phi_0(x)$. Since the Lax pair commutes $ [\hat{L},\hat{B}]=0 $ for the stationary solution, we consider the simultaneous eigenfunction:
	\begin{align}
		\hat{L}f_0=\lambda f_0,\quad \hat{B}f_0 = \omega f_0. \label{eq:eigenLB}
	\end{align}
	We first determine $ \phi_0(x) $. The first integral for the stationary SG equation is
	\begin{align}
		\phi_{0x}^2+2\cos\phi_0=\frac{4}{m}-2 \quad \leftrightarrow \quad \frac{(\frac{1}{2}\phi_{0x})^2}{1-m\sin^2\frac{\phi_0-\pi}{2}}=\frac{1}{m}.
	\end{align} 
	Integrating this equation once again yields the soliton lattice solution as
	\begin{align}
		\phi_0(x)&=\pi+ 2\operatorname{am}\left( \tfrac{x}{\sqrt{m}} \big|m \right), \label{eq:phi0} \\
		\phi_0'(x)&=\frac{2}{\sqrt{m}}\operatorname{dn}\left( \tfrac{x}{\sqrt{m}} \big|m \right),\quad \cos\tfrac{\phi_0(x)}{2}=-\operatorname{sn}\left( \tfrac{x}{\sqrt{m}} \big|m \right),\quad \sin\tfrac{\phi_0(x)}{2}=\operatorname{cn}\left( \tfrac{x}{\sqrt{m}} \big|m \right). \label{eq:phi0cossin}
	\end{align}
	Here and henceforth, we use the notations of elliptic functions in the Abramowitz-Stegun book \cite{AbramowitzStegun} unless otherwise noted. If $ 0<m<1 $, it is the rotating solution. If  $ m=1 $, it represents the stationary one-kink solution, and if $ m>1 $, it is an oscillating solution. In this paper we only consider the rotating background $ 0<m<1 $. Generally, the pair of commuting differential operators satisfies an algebraic relation $ P(\hat{L},\hat{B})=0 $ \cite{Krichever1977}. Strictly speaking, Krichever's original paper \cite{Krichever1977} only consider the case where the highest-order coefficient matrices of the matrix differential operators are invertible, but similar nature can be seen even if this assumption is not satisfied and an algebraic curve can be defined in many cases. In the present case,  $ P(\lambda,\omega)=0 $ gives a genus-one curve, i.e., an elliptic curve: 
	\begin{align}
		\lambda^2\omega^2=\lambda^4+2\left( \frac{2}{m}-1 \right)\lambda^2+1. \label{eq:ellcurve}
	\end{align}
	It can be parametrized by elliptic functions:
	\begin{align}
		\lambda(z)&=\frac{-\mathrm{i}\sqrt{m}\operatorname{sn}(\mathrm{i}z|m)\operatorname{cn}(\mathrm{i}z|m)}{\operatorname{dn}(\mathrm{i}z|m)}=\mathrm{i}\sqrt{m}\operatorname{sn}(\mathrm{i}z|m)\operatorname{sn}(\mathrm{i}z-K|m), \label{eq:lambdaz} \\
		\omega(z)&=\frac{\operatorname{sn}^2(\mathrm{i}z|m)-\operatorname{sn}^2(\mathrm{i}z-K|m)}{\lambda(z)}, \label{eq:omegaz}
	\end{align}
	where $ K=K(m) $ and $ K'=K(1-m) $. 
	The eigenfunction $ f_0(x,z) $ for the above parametrized eigenvalue $\lambda=\lambda(z)$ and the corresponding crystal momentum $k(z)$ is 
	\begin{align}
		k(z)&=\frac{Z(2\mathrm{i}z+\mathrm{i}K'|m)+Z(2\mathrm{i}z-\mathrm{i}K'|m)}{4\mathrm{i}\sqrt{m}}, \label{eq:kz} \\
		f_0(x,z)&=\frac{\mathrm{e}^{\mathrm{i}k(z)x}\Theta_4(0)}{2\Theta_4(\frac{x}{\sqrt{m}})}\begin{pmatrix} \Theta_1(\frac{x}{\sqrt{m}}-\mathrm{i}z)/\Theta_4(\mathrm{i}z) \\ \Theta_2(\frac{x}{\sqrt{m}}-\mathrm{i}z)/\Theta_3(\mathrm{i}z) \\ -\mathrm{i}\Theta_4(\frac{x}{\sqrt{m}}-\mathrm{i}z)/\Theta_1(\mathrm{i}z) \\ -\mathrm{i}\Theta_3(\frac{x}{\sqrt{m}}-\mathrm{i}z)/\Theta_2(\mathrm{i}z)  \end{pmatrix}. \label{eq:f0xz}
	\end{align}
	Here, we define the scaled theta functions by $ \Theta_i(u|m):=[\vartheta_i(\frac{\pi u}{2K},q)]_{\text{Abramowitz-Stegun}} $ with the nome $ q=\mathrm{e}^{-\pi K'/K} $, and the second argument $ m $ is omitted. Using this scaled theta function, the Jacobi zeta function is expressed as $ Z(u|m)=\frac{\mathrm{d} }{\mathrm{d} u} \ln \Theta_4(u|m) $. The derivation of Eqs.~(\ref{eq:kz}) and (\ref{eq:f0xz}) are given in \ref{app:BAfunction}. \\
	\indent $ \lambda(z),\omega(z),k(z) $ satisfy the following (quasi-)periodicity, parity, and the complex conjugation relations:
	\begin{align}
		\lambda(z)&=(-1)^{l+n}\lambda(z+nK'+\mathrm{i}lK)^{(-1)^n}=-\lambda(-z)=\lambda(z^*)^*, \label{eq:lambdasymm} \\
		\omega(z)&=(-1)^n \omega(z+nK'+\mathrm{i}lK)=-\omega(-z)=\omega(z^*)^*, \\
		k(z)&=k(z+nK'+\mathrm{i}lK)+\frac{n\pi}{2K\sqrt{m}}=-k(-z)=k(z^*)^,
	\end{align}
	for $ n,l\in\mathbb{Z} $. 
	The eigenfunction $ f_0(x,z) $ also has the symmetries
	\begin{align}
		f_0(x,z+nK'+\mathrm{i}lK)&=\mathrm{i}^l\left[ (\sigma_3^l\sigma_2^n)\otimes(\sigma_2^l\sigma_3^n) \right]f_0(x,z), \label{eq:f0periodicity} \\
		f_0(x,-z^*)&=(\sigma_3\otimes I_2)f_0(x,z)^*. \label{eq:f0ccrelation}
	\end{align}
	The two linearly independent solution for a given eigenvalue $ \lambda=\lambda(z) $ are $ f_0(x,z) $ and $ f_0(x,-z-\mathrm{i}K) $, because $ \lambda(z)=\lambda(-z-\mathrm{i}K) $.\\
	\indent To cover all eigenfunctions, the region we must consider is the rectangle with vertices $ z=\pm K'\pm \mathrm{i}K $, which is the fundamental period parallelogram of $ \lambda(z) $. The bounded eigenfunctions with $ k(z)\in\mathbb{R} $, i.e., the scattering eigenstates, lie on the lines $ z\in  \mathbb{R}+\frac{\mathrm{i}K}{2}\mathbb{Z} $, and there are four independent lines in one fundamental period parallelogram. Among them, $ \mathbb{R},\ \mathbb{R}-\mathrm{i}K $ gives monotonically increasing $ k(z) $ and real $ \lambda(z) $, while  $ \mathbb{R}\pm\frac{\mathrm{i}K}{2} $ gives monotonically decreasing $ k(z) $ and pure imaginary $ \lambda(z) $. The region satisfying $ \operatorname{Im}k(z)>0 $ is given by $ -K<\operatorname{Im}z<-\frac{K}{2},\ 0<\operatorname{Im}z<\frac{K}{2} $, and the positions of zeros of $ a(z) $ corresponding to bound eigenstates are chosen from this region in Sec.~\ref{sec:glm}.\\ %In the later section formulating the ISM, the zeros of $ a(z) $ are located in this region.\\
	\indent The completeness relation satisfied for scattering eigenstates is
	\begin{align}
		\delta(x-y)I_4 = \oint_{C_1+C_2}\frac{dz}{2\pi} f_0(x,z)f_0(y,z^*)^\dagger\sigma. \label{eq:completeness}
	\end{align}
	Here, $ C_1 $ is the rectangular contour passing through the vertices $ -K' \to K' \to K'+\frac{\mathrm{i}K}{2} \to -K'+\frac{\mathrm{i}K}{2} \to -K'  $ and $ C_2 $ is the one  $ -K'-\mathrm{i}K \to K'-\mathrm{i}K \to K'-\frac{\mathrm{i}K}{2} \to -K'-\frac{\mathrm{i}K}{2} \to -K'-\mathrm{i}K $.  Note that the integrations along vertical lines cancel out due to the periodicity. Therefore, the actual path with non-zero contribution is $ \oint_{C_1+C_2} = \int_{-K'}^{K'}-\int_{-K'+\frac{\mathrm{i}K}{2}}^{K'+\frac{\mathrm{i}K}{2}}+\int_{-K'-\mathrm{i}K}^{K'-\mathrm{i}K}-\int_{-K'-\frac{\mathrm{i}K}{2}}^{K'-\frac{\mathrm{i}K}{2}} $. The derivation of Eq.~(\ref{eq:completeness}) is given in \ref{app:completeness}.
\section{Scattering matrix}\label{sec:scatmat}
	Henceforth, we consider the scattering problem under the condition that the background potential tends to the stationary soliton lattice at spatial infinities $ x\to\pm\infty $. Let the potential $ \phi(x) $ satisfy the boundary condition
	\begin{align}
		\phi(x) \to \begin{cases} \phi_0(x) & (x\to-\infty), \\ \phi_0(x-x_0) & (x \to +\infty), \end{cases}
	\end{align}
	where $ x_0 $ represents the phase shift of the background lattice caused by solitons and ripple waves. We define the left and right Jost solutions by the asymptotic form
	\begin{align}
		f_-(x,z) &\to f_0(x,z) \quad (x \to -\infty), \\
		f_+(x,z) &\to f_0(x-x_0,z) \quad (x\to+\infty).
	\end{align}
	The scattering matrix  $ S(z) $ is then defined by the linear transformation between these two:
	\begin{align}
		\begin{pmatrix} f_+(x,z) & f_+(x,-z-\mathrm{i}K) \end{pmatrix} = \begin{pmatrix} f_-(x,z) & f_-(x,-z-\mathrm{i}K) \end{pmatrix}S(z). \label{eq:scatteringmatrix}
	\end{align}
	The uniqueness of the definition of the Jost solution by its asymptotic form and the properties of eigenfunctions (\ref{eq:f0periodicity}) and (\ref{eq:f0ccrelation}) imply
	\begin{align}
		f_\pm(x,z)=(-\mathrm{i}\sigma_3\otimes \sigma_2)^lf_\pm(x,z+2nK'+\mathrm{i}lK)=(\sigma_3\otimes I_2)f_\pm(x,-z^*)^*,\quad n,l \in \mathbb{Z}. \label{eq:fpmperiodicity}
	\end{align}
	Note that while $ f_0(x,z) $ has an additional symmetry with respect to translation of $ z $ by $ (2n+1)K' $ (see Eq.~(\ref{eq:f0periodicity})),  the corresponding symmetry does not exist for Jost solutions $ f_\pm(x,z) $, because this additional symmetry is coming from special nature of $ \phi_0(x) $ satisfying $ \hat{L}=\tilde{L} $ in Eqs.~(\ref{eq:Linv1})-(\ref{eq:Linv3}). Combining Eqs.~(\ref{eq:scatteringmatrix}) and (\ref{eq:fpmperiodicity}), we have 
	\begin{align}
		S(z)=\sigma_3^l S(z+2nK'+\mathrm{i}lK)\sigma_3^l=S(-z^*)^*=\sigma_1 S(-z-\mathrm{i}K)\sigma_1,\quad n,l \in \mathbb{Z}.
	\end{align}
	Equating the integration constant obtained by integrating Eq.~(\ref{eq:constJ}) at $ x\to\pm\infty $, we find 
	\begin{align}
		S(z)^{-1}=S(z^*)^\dagger.
	\end{align}
	We can also show $ \det S=1 $ from the constancy of the Wronskian $ \det (g_1,g_2) $ (see the sentence after Eq.~(\ref{eq:constJ})). Therefore, $S$ and its inverse matrix has the forms
	\begin{align}
		S(z) &= \begin{pmatrix} a(z) & -b(z^*)^* \\ b(z) & a(z^*)^*  \end{pmatrix}, \\
		S(z)^{-1}&=\begin{pmatrix} a(z^*)^* & b(z^*)^* \\ -b(z) & a(z)  \end{pmatrix},
	\end{align}
	with matrix elements satisfying
	\begin{align}
		&a(z)a(z^*)^*+b(z)b(z^*)^*=1,\\ 
		&a(z)=a(-z^*)^*=a(z-\mathrm{i}K),\quad b(z)=b(-z^*)^*=-b(z-\mathrm{i}K). \label{eq:scatinvo2}
	\end{align}
	From this, the zeros of $ a(z) $ always appear simultaneously at four points $ z,-z^*, z-\mathrm{i}K, -z^*-\mathrm{i}K $, and the orders of these zeros are the same.
\section{Integral representation of Jost solution and Gelfand-Levitan-Marchenko equation}\label{sec:glm}
	Let us introduce the integral representation for the left Jost solution
	\begin{align}
		f_-(x,z)=f_0(x,z)+\int_{-\infty}^x dy \Gamma(x,y)f_0(y,z). \label{eq:kernelrep}
	\end{align}
	The kernel function $ \Gamma(x,y) $ is a $4\times 4$ matrix and assumed to decrease exponentially in the limits $ x,y \to -\infty $.
	In order for this integral to converge, the integrand must decrease $ y\to -\infty $, and this condition is satisfied at least for eigenfunctions with $ \operatorname{Im}k \le 0 $, i.e.,  $ -K/2 \le  \operatorname{Im}z \le 0,\ K/2 \le \operatorname{Im}z \le K  $. From Eq.~(\ref{eq:invol}), we immediately conclude that
	 \begin{align}
	 	\Gamma(x,y)=(\sigma_3\otimes\sigma_2)\Gamma(x,y)(\sigma_3\otimes\sigma_2)=(\sigma_3\otimes I_2)\Gamma(x,y)^*(\sigma_3\otimes I_2)=(I_2\otimes\sigma_2)\Gamma(x,y)^*(I_2\otimes\sigma_2). \label{eq:gammainvol}
	 \end{align}
	Let us now write
	 \begin{align}
	 	\hat{L}_0(x)&=-4\mathrm{i}\sigma_3\oplus O_2+U_0(x),\quad U_0(x)=\begin{pmatrix} \mathrm{i}\sigma_1 w_0 & \mathrm{e}^{-\mathrm{i}\phi_0\sigma_2/2} \\ \mathrm{e}^{-\mathrm{i}\phi_0\sigma_2/2} & O_2 \end{pmatrix}, \\
	 	\hat{L}(x)&=-4\mathrm{i}\sigma_3\oplus O_2+U(x),\quad U_0(x)=\begin{pmatrix} \mathrm{i}\sigma_1 w & \mathrm{e}^{-\mathrm{i}\phi\sigma_2/2} \\ \mathrm{e}^{-\mathrm{i}\phi\sigma_2/2} & O_2 \end{pmatrix}.
	 \end{align}
	Following the same derivation as Ref.~\cite{doi:10.1093/ptep/ptw020}, section 2.8, (In the present case, the first-order coefficient matrix of $ \hat{L} $ is not full-rank, but it makes no difference to the proof.)
	\begin{gather}
		U(x)-U_0(x)=4\mathrm{i}\big[\sigma_3\oplus O_2, \Gamma(x,x)\big], \label{eq:kerneltopotential} \\
		\hat{L}(x)\Gamma(x,y)=\Gamma(x,y)\hat{L}_0(y).
	\end{gather}
	From Eq.~(\ref{eq:kerneltopotential}),
	\begin{align}
		w&=w_0+2\Gamma_{12}, \\
		\mathrm{e}^{\mathrm{i}\phi/2}&=\mathrm{e}^{\mathrm{i}\phi_0/2}+4(\mathrm{i}\Gamma_{13}+\Gamma_{14}). \quad \left(\leftrightarrow \cos\tfrac{\phi}{2}=\cos\tfrac{\phi_0}{2}+4\mathrm{i}\Gamma_{13},\ \sin\tfrac{\phi}{2}=\sin\tfrac{\phi_0}{2}-4\mathrm{i}\Gamma_{14}.\right), \label{eq:phifromGamma}
	\end{align}
	where we briefly write the matrix element of $ \Gamma(x,x) $ as $ \Gamma_{ij}=[\Gamma(x,x)]_{ij} $. 
	Note that $ \Gamma_{12},\ \mathrm{i}\Gamma_{13} $, and $ \mathrm{i}\Gamma_{14} $ are real-valued functions due to Eq.~(\ref{eq:gammainvol}). Using these relations, we can re-construct the potential from the kernel $ \Gamma $, which is determined from scattering data by solving the GLM equation shown below.\\
	\indent The integral kernel $ \Gamma(x,y) $ is uniquely determined from the following scattering data: (i) the values of the reflection coefficient for scattering states $ r(z)=b(z)/a(z) $, where scattering states mean the bounded eigenfunction s.t. $ k(z)\in\mathbb{R} $, appearing on lines  $ z\in\mathbb{R}+\frac{\mathrm{i}K}{2}\mathbb{Z} $, and (ii) the list of the zeros of $ a(z) $, which we write $ z_1,\dots,z_n $, and the normalization factor $ c_j^2:=\frac{b(z_j)}{\mathrm{i}\dot{a}(z_j)} $, where $ z_j $'s are to be chosen from the region $ \operatorname{Im}k(z_j)>0 $, and due to the property (\ref{eq:scatinvo2}), the zeros simultaneously appear at $ z,-z^*, z-\mathrm{i}K, -\mathrm{z}^*-\mathrm{i}K $. The values of $ c_j^2 $ also have some constraint, whose detail will be described in Sec.~\ref{sec:realcondition}. Here, we only treat the case where  all zeros of $ a(z) $ are first order. For uniform background, the solitons corresponding to higher-order zeros are discussed in Ref.~\cite{doi:10.1143/JPSJ.53.2908}. The case of unstable uniform background is considered in Ref.~\cite{19873069}. \\
	\indent The GLM equation that determines the kernel $ \Gamma(x,y) $ from the above-mentioned scattering data is given by
	\begin{align}
		\Gamma(x,y)+\Omega(x,y)+\int_{-\infty}^x\mathrm{d}w\Gamma(x,w)\Omega(w,y)=0,\quad y\le x. \label{eq:glm}
	\end{align}
	Here, $ \Omega=\Omega_{\text{bd}}+\Omega_{\text{sc}} $ with
	\begin{align}
%		\Omega_{\text{bd}}(x,w)&=\sum_{j=1}^n f_0(x,-z_j-\mathrm{i}K)c_j^2f_0(w,z_j^*)^\dagger\sigma, \\
		\Omega_{\text{bd}}(x,y)&:=W(x)W(y)^T(I_2\otimes \sigma_1), \label{eq:omegabd} \\
		W(x)&:=-\mathrm{i}\sigma_3\otimes \sigma_2 \left( f_0(x,-z_1)c_1,\dots, f_0(x,-z_n)c_n \right), \label{eq:omegabd2} \\
		\Omega_{\text{sc}}(x,y)&:=\left[ \int_{-K'}^{K'}-\int_{-K'+\frac{\mathrm{i}K}{2}}^{K'+\frac{\mathrm{i}K}{2}}+\int_{-K'-\mathrm{i}K}^{K'+\mathrm{i}K}-\int_{-K'-\frac{\mathrm{i}K}{2}}^{K'-\frac{\mathrm{i}K}{2}} \right]\frac{\mathrm{d}z}{2\pi} f_0(x,-z-\mathrm{i}K)r(z)f_0(y,z^*)^\dagger\sigma. \label{eq:omegasc}
	\end{align}
	 $ \Omega_{\text{bd}} $ represents the contribution from bound states, and $ \Omega_{\text{sc}} $ is that from scattering states. The derivation of the GLM equation (\ref{eq:glm}) with (\ref{eq:omegabd})-(\ref{eq:omegasc}) is given in \ref{app:glm}.
\section{Reflectionless potentials}\label{sec:reflectionlesspot}
	\indent Let us solve the GLM equation (\ref{eq:glm}) for the reflectionless case, i.e., $ \Omega_{\text{sc}}=0 $. The ansatz for the kernel $ \Gamma(x,y) $ is as follows. Let us introduce the notation $ H(x)=(h_1(x),\dots,h_n(x)) $, where each $ h_i(x) $ is four-component column vector. We then assume
	\begin{align}
		\Gamma(x,y)=H(x)W(y)^T(I_2\otimes \sigma_1).
	\end{align}
	Substituting it to the GLM equation, we obtain the equation for $ H(x) $: 
	\begin{align}
		H(x)+W(x)+H(x)G(x)=0, \label{eq:eqforH}
	\end{align}
	where we define the $ n\times n $ Gram matrix by
	\begin{align}
		G(x)=\int_{-\infty}^x dy W(y)^T(I_2\otimes \sigma_1)W(y).
	\end{align}
	From Eq.~(\ref{eq:eqforH}),  $ H=-W(I_n+G)^{-1} $.  Let  $ G_{lj}(x) $ be the $ (l,j) $-component of $ G(x) $. Using Eq.~(\ref{eq:constJ}) and the symmetries of $ \lambda(z) $ and $ f_0(x,z) $, Eqs.~(\ref{eq:lambdasymm}), (\ref{eq:f0periodicity}), and (\ref{eq:f0ccrelation}), we find
	\begin{align}
		G_{lj}(x) = c_lc_j\frac{4 f_0(x,-z_l)^T(\sigma_2\oplus O_2)f_0(x,-z_j)}{\lambda(z_l)-\lambda(z_j)}.
	\end{align}
	It can be simplified using the Weierstrass three-term formula in Ref.~\cite{Kharchev201519}, Eq.~(3.8). The result is
	\begin{align}
		G_{lj}(x)=c_lc_j\mathrm{e}^{-\mathrm{i}[k(z_l)+k(z_j)]x}\frac{\Theta_2\Theta_4\Theta_3(\frac{x}{\sqrt{m}}+\mathrm{i}(z_l+z_j))}{\Theta_3\Theta_2(\mathrm{i}(z_l+z_j))\Theta_4(\frac{x}{\sqrt{m}})}.
	\end{align}
	Thus,
	\begin{align}
		\Gamma(x,y)=-W(x)[I_n+G(x)]^{-1}W(y)^T(I_2\otimes \sigma_1).
	\end{align}
	Let us write $ W $ as an array of row vectors: $ W=\left( \begin{smallmatrix} W_1 \\ W_2 \\ W_3 \\ W_4 \end{smallmatrix} \right) $, where each $ W_i $ is a $ 1\times n $ row vector. Then we get the expressions $ \Gamma_{13}=-W_1(I_n+G)^{-1}W_4^T,\ \Gamma_{14} = -W_1(I_n+G)^{-1} W_3^T $. Using the relation (\ref{eq:phifromGamma}) and the Woodbury-type identity $ 1+\vec{a}^\dagger A^{-1} \vec{b}=\frac{\det(A+\vec{b}\vec{a}^\dagger)}{\det A}  $, 
	\begin{align}
		\cos\tfrac{\phi}{2}&=\cos\tfrac{\phi_0}{2}+4\mathrm{i}\Gamma_{13}=\left(\cos\tfrac{\phi_0}{2}\right)\frac{\det(I+P)}{\det(I+G)},\quad P_{ij}=G_{ij}-\frac{4\mathrm{i}}{\cos\frac{\phi_0}{2}}f_{03}(x,-z_i)f_{02}(x,-z_j), \\
		\sin\tfrac{\phi}{2}&=\sin\tfrac{\phi_0}{2}-4\mathrm{i}\Gamma_{14}=\left(\sin\tfrac{\phi_0}{2}\right)\frac{\det(I+Q)}{\det(I+G)},\quad Q_{ij}=G_{ij}-\frac{4\mathrm{i}}{\sin\frac{\phi_0}{2}}f_{04}(x,-z_i)f_{02}(x,-z_j), 
	\end{align}
	where $ f_{0j}(x,z) $ represents the $ j $-th component of $ f_0(x,z) $ defined by (\ref{eq:f0xz}). These can be simplified by using Ref.~\cite{Kharchev201519}, Eq.~(3.5b). Rewriting $ C_j=\frac{1}{c_j^2\sqrt{m}} $, and introducing new matrices $ \mathcal{G},\mathcal{P},\mathcal{Q} $ through the relations $ G_{ij}=m^{1/4}c_ic_j\mathrm{e}^{-\mathrm{i}[k(z_i)+k(z_j)]x}\mathcal{G}_{ij},\ P_{ij}=m^{1/4}c_ic_j\mathrm{e}^{-\mathrm{i}[k(z_i)+k(z_j)]x}\mathcal{P}_{ij},\ Q_{ij}=m^{1/4}c_ic_j\mathrm{e}^{-\mathrm{i}[k(z_i)+k(z_j)]x}\mathcal{Q}_{ij} $, we obtain the final expressions
	\begin{align}
		\cos\tfrac{\phi}{2}&=\left( \cos\tfrac{\phi_0}{2} \right)\frac{\det(\mathcal{E}+\mathcal{P})}{\det(\mathcal{E}+\mathcal{G})}, \label{eq:nsolC} \\
		\sin\tfrac{\phi}{2}&=\left( \sin\tfrac{\phi_0}{2} \right)\frac{\det(\mathcal{E}+\mathcal{Q})}{\det(\mathcal{E}+\mathcal{G})}, \label{eq:nsolS}
	\end{align}
	where the components of  $ n\times n $ matrices $ \mathcal{E},\ \mathcal{G},\ \mathcal{P},\ $ and $ \mathcal{Q} $ are given by
	\begin{align}
		\mathcal{E}_{ij}&=\delta_{ij}C_j\mathrm{e}^{2\mathrm{i}k(z_j)x}, \label{eq:calEij} \\
		\mathcal{G}_{ij}&=\frac{\Theta_4 \Theta_3(\frac{x}{\sqrt{m}}+\mathrm{i}(z_i+z_j))}{\Theta_2(\mathrm{i}(z_i+z_j))\Theta_4(\frac{x}{\sqrt{m}})}, \label{eq:calGij} \\
		\mathcal{P}_{ij}&=\left( \frac{\Theta_4(\mathrm{i}z_i)}{\Theta_1(\mathrm{i}z_i)} \right) \left( \frac{-\Theta_4\Theta_2(\frac{x}{\sqrt{m}}+\mathrm{i}(z_i+z_j))}{\Theta_2(\mathrm{i}(z_i+z_j))\Theta_1(\frac{x}{\sqrt{m}})} \right) \left( \frac{\Theta_2(\mathrm{i}z_j)}{\Theta_3(\mathrm{i}z_j)} \right),\\
		\mathcal{Q}_{ij}&= \left( \frac{\Theta_4(\mathrm{i}z_i)}{\Theta_2(\mathrm{i}z_i)} \right) \left( \frac{\Theta_4\Theta_1(\frac{x}{\sqrt{m}}+\mathrm{i}(z_i+z_j))}{\Theta_2(\mathrm{i}(z_i+z_j))\Theta_2(\frac{x}{\sqrt{m}})} \right) \left( \frac{\Theta_1(\mathrm{i}z_j)}{\Theta_3(\mathrm{i}z_j)} \right). \label{eq:calQij}
	\end{align}
	Since the wavenumbers used to make bound states all satisfy $ \operatorname{Im}k(z_j) >0 $, 
	\begin{align}
		\mathcal{E} \to \begin{cases} \infty &  (x \to -\infty), \\ 0 & (x \to +\infty). \end{cases} 
	\end{align}
	Therefore, the asymptotic form of $ \phi $ is given by
	\begin{align}
		\cos\tfrac{\phi}{2} &\to \begin{cases} \cos\tfrac{\phi_0}{2} & (x\to -\infty), \\ \cos\tfrac{\phi_0}{2}\det \mathcal{P}\mathcal{G}^{-1} & (x\to+\infty), \end{cases}\\
		\sin\tfrac{\phi}{2} &\to \begin{cases} \sin\tfrac{\phi_0}{2} & (x\to -\infty), \\ \sin\tfrac{\phi_0}{2}\det \mathcal{Q}\mathcal{G}^{-1} & (x\to+\infty). \end{cases}
	\end{align}
	The determinant appearing in $ x\to+\infty $ is calculated in \ref{app:dettheta}. The resultant asymptotic form is 
	\begin{align}
		&\phi(x) \to \begin{cases} \phi(x) & (x\to -\infty), \\ \phi(x-x_0) & (x\to+\infty),  \end{cases} \label{eq:x0asymp0} \\
		&x_0:=-2\mathrm{i}\sqrt{m}\sum_{j=1}^nz_n. \label{eq:x0asymp}
	\end{align}
	 $ x_0 $ has the meaning of the phase shift of the background lattice caused by solitons. The realness of $ x_0 $ follows from the fact that the zeros of $ a(z) $ always simultaneously emerge at four points $ z,z-\mathrm{i}K,-z^*, $ and $ -z^*-\mathrm{i}K $.
\section{Time evolution}\label{sec:timeevo}
	Finally, let us determine the time dependence of scattering matrix if the system obeys the Lax equation (\ref{eq:lax01}). Let $ f $ be a time-dependent eigenfunction of $ \hat{L} $. Then, the Lax equation implies 
	\begin{align}
		\hat{L}f &= \lambda f, \\
		4\mathrm{i}f_t&=-\hat{B}f.
	\end{align}
	We now define the time-dependent eigenfunction by the initial condition $ \tilde{f}_+(t=0,x,z)=f_+(x,z) $. Though $ \hat{B} $ is now a time-dependent operator for finite $ x $, it is time-independent at $ x\to\pm\infty $. Therefore, the asymptotic forms of this $ \tilde{f}_+(t,x,z) $ is given by  
	\begin{align}
		(\tilde{f}_+(t,x,z),\tilde{f}_+(t,x,-z-\mathrm{i}K)) \to \begin{cases} (f_0(x,z),f_0(x,-z-\mathrm{i}K))\mathrm{e}^{\mathrm{i}\omega(z)t\sigma_3/4}S(z) & ( x\to -\infty), \\ (f_0(x-x_0,z),f_0(x-x_0,-z-\mathrm{i}K))\mathrm{e}^{\mathrm{i}\omega(z)t\sigma_3/4} & (x\to+\infty).  \end{cases}
	\end{align}
	Here, we have used the relation $ \omega(-z-\mathrm{i}K)=-\omega(z) $. Therefore, if we define the time-dependent right Jost solution by the asymptotic form $ f_+(t,x,z) \to f_0(x-x_0,z) \text{ for } x\to+\infty $, which is different from $ \tilde{f}_+(t,x,z) $, then the relation $ (\tilde{f}_+(t,x,z),\tilde{f}_+(t,x,-z-\mathrm{i}K)) = (f_+(t,x,z),f_+(t,x,-z-\mathrm{i}K))\mathrm{e}^{\mathrm{i}\omega(z)t/4} $ holds. Then the time evolution of the scattering matrix is given by
	\begin{align}
		S(t,z)=\mathrm{e}^{\mathrm{i}\omega(z)t\sigma_3/4}S(z)\mathrm{e}^{-\mathrm{i}\omega(z)t\sigma_3/4},
	\end{align}
	or for each component,
	\begin{align}
		a(t,z)=a(z),\quad b(t,z)=\mathrm{e}^{-\mathrm{i}\omega(z)t/2}b(z).
	\end{align}
	Since $ C_j $ is defined by $ C_j= \frac{1}{c_j^2\sqrt{m}}=\frac{\mathrm{i}\dot{a}(z_j)}{\sqrt{m}b(z_j)} $, its time evolution is written as
	\begin{align}
		C_j(t)=C_j\mathrm{e}^{\mathrm{i}\omega(z_j)t/2}.
	\end{align}
	Therefore, we can define the time-dependent $ \mathcal{E}_{ij}(t) $ as follows. In Eq.~(\ref{eq:calEij}), we replace $ C_j $ by $ C_j(t)=C_j\mathrm{e}^{\mathrm{i}\omega(z_j)t/2} $, i.e.,
	\begin{align}
		\mathcal{E}_{ij}(t)=\delta_{ij}C_j\mathrm{e}^{2\mathrm{i}k(z_j)x+\mathrm{i}\omega(z_j)t/2}.
	\end{align}
	Then the time-dependent solution of the SG equation is given by replacing $ \mathcal{E} $ with $ \mathcal{E}(t) $ in Eqs. (\ref{eq:nsolC}) and (\ref{eq:nsolS}):
		\begin{align}
		\cos\tfrac{\phi}{2}&=\left( \cos\tfrac{\phi_0}{2} \right)\frac{\det(\mathcal{E}(t)+\mathcal{P})}{\det(\mathcal{E}(t)+\mathcal{G})}, \label{eq:tdcos} \\
		\sin\tfrac{\phi}{2}&=\left( \sin\tfrac{\phi_0}{2} \right)\frac{\det(\mathcal{E}(t)+\mathcal{Q})}{\det(\mathcal{E}(t)+\mathcal{G})}, \label{eq:tdsin}
	\end{align}
	where the definition of $ \mathcal{G}, \mathcal{P} $, and $ \mathcal{Q} $ remains unchanged (Eqs.~(\ref{eq:calGij})-(\ref{eq:calQij})).
\section{Limitation to eigenvalues $ z_j $'s and normalization coefficients $ C_j $'s}\label{sec:realcondition}
	\indent In order for the function $ \phi(x,t) $ to be real and bounded, $ z_j $'s, which are the zeros of $ a(z) $ corresponding to the discrete eigenvalues for bound states, and the normalization coefficients $ C_j $'s must satisfy several conditions.\footnote{If we consider application to scientific problems where the complex-valued and/or divergent solutions have some appropriate physical interpretations, the restriction stated here can be loosened.} Here we describe it. \\
\indent First, depending on its value, $ z_j $'s must satisfy the following (i) or (ii):
	\begin{enumerate}[(i)]
		\item  $ z_j,\ z_j-\mathrm{i}K,\ -z_j^*,\ -z_j^*-\mathrm{i}K $ appear simultaneously. Due to this symmetry, one of these four $ z_j $ can be chosen from $ 0 \le \operatorname{Re}z_j\le K',\ 0< \operatorname{Im}z_j<\frac{K}{2} $ without loss of generality.
		\item If, as a special case, $ z_j= \mathrm{i}r_j $ or $z_j= \mathrm{i}r_j+K'  $ with $ r_j\in (0,\frac{K}{2}), $ then just two zeros $ z_j $ and $ z_j-\mathrm{i}K $ appear simultaneously. 
	\end{enumerate}
	The case (i) corresponds to the breather solution and the case (ii) the traveling one kink solution.
	The total number of  $ z_j $'s are always even. 
	If there are breathers and even number of kinks, the number is a multiple of four, while if there exists odd number of kinks, the number is of the form $4n+2$. The velocity of the soliton is given by  $ v_j=-\frac{\operatorname{Im}\omega(z_j)}{4\operatorname{Im}k(z_j)} $. In particular, for one kink solution, it reduces to $ v_j=-\frac{\omega(z_j)}{4k(z_j)} $, which is negative if $ z_j=\mathrm{i}r_j $ and positive if $ z_j=\mathrm{i}r_j+K' $. \\
	\indent Next let us consider the condition for $ C_j $'s. By Eq.~(\ref{eq:scatinvo2}), the derivative of $ a(z) $ satisfies
	\begin{align}
		\dot{a}(z)=-\dot{a}(-z^*)^*=\dot{a}(z-\mathrm{i}K),
	\end{align}
	and thus if we define
	\begin{align}
		C(z):=\frac{\mathrm{i}\dot{a}(z)}{\sqrt{m}b(z)},
	\end{align}
	then it has the symmetry
	\begin{align}
		C(z)=C(-z^*)^*=-C(z-\mathrm{i}K).
	\end{align}
	This symmetry is the same as that of $ b(z) $ given in Eq.~(\ref{eq:scatinvo2}). Therefore, for the breather solution where four zeros appear, if we temporarily write these four as $ z_j, z_{j+1}=z_j-\mathrm{i}K,\ z_{j+2}=-z_j^*,\ z_{j+3}=-z_j^*-\mathrm{i}K $, then we should choose the corresponding coefficients as $ C_{j+1}=-C_j,\ C_{j+2}=C_j^{*},\ C_{j+3}=-C_j^{*} $. For the kink solution, where two zeros $ z_j $ and $ z_{j+1}=z_j-\mathrm{i}K $ appear, $ C_{j+1}=-C_j $ are both real and must have the opposite sign. Whether the solution becomes kink or anti-kink, i.e., the phase rotation becomes counterclockwise or clockwise, is determined by which one is chosen positive.
\section{Animation of soliton solutions}\label{sec:anime}
\begin{figure}[tb]
	\begin{center}
	\includegraphics[width=.7\textwidth]{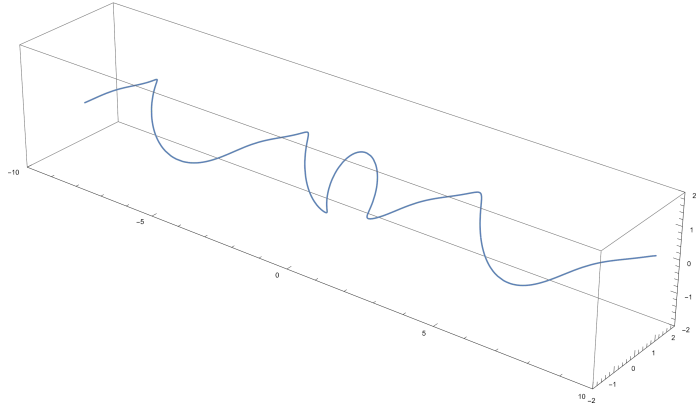}
	\caption{\label{fig:onesolsnapshot} A snapshot of one kink solution. The number of zeros is $ n=2 $ and the parameters are $ m=0.95,\ z_1=0.02\mathrm{i}K+K',\ z_2=z_1-\mathrm{i}K,\ C_1=-1,\ C_2=-C_1 $. The snapshot at $ t=1 $.}
	\end{center}
\end{figure}

Here, we present a few examples of the soliton solutions, Eqs.~(\ref{eq:tdcos}) and (\ref{eq:tdsin}), by gif animations. We visualize the solution through the 3D plot $ (x,\, \cos\phi(x,t),\, \sin\phi(x,t)) $. An example of snapshot is shown in Fig.~\ref{fig:onesolsnapshot}.
 See attached files, testx.gif with x=1,2, and 3. Below, we provide the parameters for each solution. We write $ K=K(m) $ and $ K'=K(1-m) $. 
\begin{enumerate}[(i)]
\item test1.gif: one kink solution. \\
The number of zeros is $ n=2 $. The parameters are $ m=0.99,\ z_1=0.02\mathrm{i}K+K',\ z_2=z_1-\mathrm{i}K,\ C_1=-1,\ C_2=-C_1 $. The time range is $ -10\le t \le 10 $. 
\item test2.gif: one anti-kink solution. \\
 $  C_1=1,\ C_2=-C_1 $, and other parameters are the same as (i). 
\item test3.gif: one breather solution. \\
The number of zeros is $ n=4 $. The parameters are $ m=0.99,\ z_1=0.1\mathrm{i}K+0.5K',\ z_2=z_1-\mathrm{i}K,\ z_3=-z_1^*,\ z_4=-z_2^*,\ C_1=-1,\ C_2=-C_1,\ C_3=C_1^*,\ C_4=C_2^* $. The time range is $ -10\le t \le 10 $.  
\end{enumerate}
Solutions with more number of solitons will be generated using the Mathematica file in google drive.\footnote{In this file, since the default build-in Jacobi elliptic functions are a little slow, the functions defined by the ratio of theta functions are used alternatively.}

\section{Summary and discussion}\label{sec:summary}
In this paper, we have derived the multi-soliton solutions of the SG equation with elliptic-function background by the ISM. The result is expressed by a determinant of theta functions, i.e., Eqs. (\ref{eq:tdcos}) and (\ref{eq:tdsin}). The shift of the background lattice caused by solitons (\ref{eq:x0asymp0}) with (\ref{eq:x0asymp}) has also been found using the addition formula of theta functions. One key tool in our work is the Lax pair represented by $ 4\times 4 $ matrix differential operators (\ref{eq:lax02}) and (\ref{eq:lax03}), originally introduced in Ref.~\cite{Takhtadzhyan1974}. This most conventional but seemingly outdated approach makes it possible to use the common form of the integral representation of the Jost solution (\ref{eq:kernelrep}) without modification and the formulation of the ISM is simplified. Also, the completeness relations in an indefinite inner product space, which is necessary in the formulation of the ISM, has been discussed in detail (\ref{app:completeness}). 
Application to various physical phenomena including the dynamics of defects with periodic background, extension to unstable oscillating background ($m>1$) and studying multi rogue waves, and the solitons associated with higher-order zeros of $ a(z) $, are left as possible future works. \\
\indent Before concluding, we provide several technical remarks and perspectives. In this work, we can introduce the concept of orthogonality between eigenfunctions based on the indefinite inner product, since the Lax operators have the symmetry (\ref{eq:sigmahermitian}). If the Lax pair has no such symmetry and any inner product cannot be defined, the completeness relation will be constructed from the set of left and right eigenfunctions, an analog of left and right eigenvectors for finite dimensional matrix. Such basis is sometimes called a bi-orthogonal basis.\\
\indent Whether we can always reduce the classical integrable systems written by zero-curvature condition (or a compatibility condition) and/or the Lax pair including an integral operator to the Lax formalism with differential operators seems to be unclear. (The inverse operation is easy; if one has an integrable equation written by the Lax form using matrix differential operators, one can immediately rewrite it in a zero-curvature expression.) If we can do it, the integral representation of the Jost solution will be widely applicable and multi-soliton solutions will be easily obtained by the dressing method \cite{ZS1974}, where we can even omit the full formulation of the ISM if we do not have an interest in general initial-value problem. We also note that, as emphasized in Secs.~\ref{sec:laxsymsigmaortho} and \ref{sec:solitonlattice}, the highest-order coefficient matrix of the Lax operator $ \hat{L} $ is not full-rank and $ \hat{L}^{-1} $ can be written down explicitly without using an integral operator. 
Though we do not show detail here, a similar property also emerges in the derivative nonlinear Schr\"odinger equation, which is usually treated by the Kaup-Newell form. That is, if we rewrite it to the Lax form, the highest-order coefficient matrix is not full-rank. 
Even in these systems, the algebraic curve from the pair of commuting operators can be appropriately defined (see Eq.~(\ref{eq:ellcurve})), though, strictly speaking, these cases are not included in the seminal work \cite{Krichever1977}. Exploring these examples more comprehensively and exhaustively might bring a slight extension to the theory of commuting differential operators.

\ack
This work was supported by MEXT-Supported Program Grant No.~S1511006 and JSPS KAKENHI Grant No.~JP19H05821.

\appendix
\section{Integral often emerging in calculation of eigenfunctions for elliptic potentials}\label{app:FAI}
	Let us define $ \operatorname{scd}(z|m):=\operatorname{sn}(z|m)\operatorname{cn}(z|m)\operatorname{dn}(z|m) $. When an elliptic-function potential is given in some hierarchy of integrable systems, the calculation of its eigenfunctions often reduces to the following integral: 
	\begin{align}
		\int dx \frac{\alpha [\operatorname{scd}(\alpha x)+ \operatorname{scd}\beta]}{\operatorname{sn}^2(\alpha x)-\operatorname{sn}^2\beta}=\alpha Z(\beta)x +\ln \frac{\Theta_1(\alpha x-\beta)}{\Theta_4(\beta)\Theta_4(\alpha x)},
	\end{align}
	even if the corresponding Riemann surface has higher genus $ g>1 $. This formula can be derived by using \cite{PhysRevE.93.062224}, appendix B. If we interpret this formula using Weierstrass's functions, it reduces to the formula given in Ref.~\cite{HurwitzCourant}, II, Kap.~6, \S3, and is called the standard form of the elliptic integral of the third kind.  Using it, the solution to the first-order differential equation
	\begin{align}
		\frac{f_x}{f}=\gamma+\frac{\alpha [\operatorname{scd}\alpha x+ \operatorname{scd}\beta]}{\operatorname{sn}^2\alpha x-\operatorname{sn}^2\beta}
	\end{align}
	is given by
	\begin{align}
		f(x)= C \mathrm{e}^{\mathrm{i}kx}\frac{\Theta_1(\alpha x- \beta)}{\Theta_4(\beta)\Theta_4(\alpha x)},\ k=-\mathrm{i}(\gamma+\alpha Z(\beta)).
	\end{align}
	If $ m\in [0,1] $, $ \alpha>0 $, and $ k $ is real, then $ f $ becomes a twisted periodic function
	\begin{align}
		f\left( x+\tfrac{4Kl}{\alpha} \right)=\mathrm{e}^{\mathrm{i}k\frac{4Kl}{\alpha}}f(x),\quad l\in\mathbb{Z}.
	\end{align}
	Therefore, if $ k $ can be physically interpreted to be crystal momentum of the Bloch function, it is defined up to $ \operatorname{mod} \frac{\alpha \pi}{2K} $ and the corresponding Brillouin zone is given by $ [-\frac{\alpha \pi}{4K},\frac{\alpha \pi}{4K}] $.
\section{Eigenfunctions for soliton lattice}\label{app:BAfunction}
	In this appendix, we derive the simultaneous eigenfunction $ f_0(x,z) $ with its crystal momentum $ k(z) $ in Eqs.~(\ref{eq:kz}) and (\ref{eq:f0xz}) for the time-independent Lax pair $ \hat{L} $ and $ \hat{B} $ with the soliton lattice potential (\ref{eq:phi0}), when the eigenvalues are parametrized by Eqs.~(\ref{eq:lambdaz}) and (\ref{eq:omegaz}). The method is in fact the special case of the one given by Krichever \cite{Krichever1977}.\\
	\indent Let us write $ f_0=\left(\begin{smallmatrix}g \\ \frac{1}{\lambda}\mathrm{e}^{-\mathrm{i}\phi\sigma_2}g\end{smallmatrix}\right), $ where $ g $ is a two-component vector $ g=\left(\begin{smallmatrix}g_1 \\ g_2 \end{smallmatrix}\right) $. Eliminating $ g_1' $ and $ g_2' $ using the eigenequation of $ \hat{L} $ in Eq.~(\ref{eq:eigenLB}), and substituting them to that of $ \hat{B} $, one obtains two linear relations with respect to $ g_1 $ and $ g_2 $. The determinant of this coefficient matrix must vanish, which gives an elliptic curve (\ref{eq:ellcurve}). Using it, one can eliminate either of $ g_1 $ and $ g_2 $ and obtain the first-order differential equation only containing one function: 
	\begin{align}
		\frac{g_1'}{g_1}&=\frac{-\mathrm{i}(2\lambda-\omega)}{4}+\frac{\mathrm{i}(-1+\lambda^4-2\lambda^3\omega+\lambda^2\omega^2)}{2\lambda(-1+\lambda^2-\lambda\omega+2\cos^2\frac{\phi_0}{2})}+\frac{-\phi_{0x}\cos\frac{\phi_0}{2}\sin\frac{\phi_0}{2}}{-1+\lambda^2-\lambda\omega+2\cos^2\frac{\phi_0}{2}},\\
		\frac{g_2'}{g_2}&=\frac{\mathrm{i}(2\lambda+\omega)}{4}+\frac{-\mathrm{i}(-1+\lambda^4+2\lambda^3\omega+\lambda^2\omega^2)}{2\lambda(-1+\lambda^2+\lambda\omega+2\cos^2\frac{\phi_0}{2})}+\frac{-\phi_{0x}\cos\frac{\phi_0}{2}\sin\frac{\phi_0}{2}}{-1+\lambda^2+\lambda\omega+2\cos^2\frac{\phi_0}{2}}.
	\end{align}
	Using Eq.~(\ref{eq:phi0cossin}) and the parametrizations (\ref{eq:lambdaz}) and (\ref{eq:omegaz}), the above equations reduce to
	\begin{align}
		\frac{g_1'}{g_1}&=\frac{-\mathrm{i}(2\lambda-\omega)}{4}+\frac{1}{\sqrt{m}}\frac{\operatorname{scd}(\frac{x}{\sqrt{m}})+\operatorname{scd}(\mathrm{i}z)}{\operatorname{sn}^2(\frac{x}{\sqrt{m}})-\operatorname{sn}^2(\mathrm{i}z)}, \\
		\frac{g_2'}{g_2}&=\frac{\mathrm{i}(2\lambda+\omega)}{4}+\frac{1}{\sqrt{m}}\frac{\operatorname{scd}(\frac{x}{\sqrt{m}})+\operatorname{scd}(\mathrm{i}z-K)}{\operatorname{sn}^2(\frac{x}{\sqrt{m}})-\operatorname{sn}^2(\mathrm{i}z-K)}.
	\end{align}
	Therefore, using the formula in \ref{app:FAI}, we find
	\begin{align}
		g_1&=C_1\mathrm{e}^{\mathrm{i}k_1x}\frac{\Theta_1(\frac{x}{\sqrt{m}}-\mathrm{i}z)}{\Theta_4(\mathrm{i}z)\Theta_4(\frac{x}{\sqrt{m}})},\ k_1=\frac{\omega-2\lambda}{4}-\frac{\mathrm{i}}{\sqrt{m}}Z(\mathrm{i}z), \label{eq:g1c1k1} \\
		g_2&=C_2\mathrm{e}^{\mathrm{i}k_2x}\frac{\Theta_2(\frac{x}{\sqrt{m}}-\mathrm{i}z)}{\Theta_3(\mathrm{i}z)\Theta_4(\frac{x}{\sqrt{m}})},\ k_2=\frac{\omega+2\lambda}{4}-\frac{\mathrm{i}}{\sqrt{m}}Z(\mathrm{i}z-K).
	\end{align}
	Due to the Bloch (Floquet) theorem, $ g_1 $ and $ g_2 $ must share the same crystal momentum, i.e., $ k_1 $ and $ k_2 $ must be related by $ k_1\equiv k_2 \mod \frac{\pi}{2\sqrt{m}K} $. In fact, using the periodicity of the Jacobi elliptic and zeta functions, we can check $ k_1=k_2 $. Therefore, henceforth we simply write $ k_1=k_2=k(z) $. The ratio $ C_1/C_2 $ must also be fixed. Rewriting $ g_1/g_2 $ using the Jacobi elliptic function, and checking its consistency with the linear relation between $ g_1 $ and $ g_2 $ obtained from the eigenequation of $ \hat{B} $, we find $ C_1/C_2=1 $. \\
	\indent The crystal momentum $ k(z) $ can be further simplified as follows. The derivative $ k'(z) $ can be simplified using the double-angle formula of the $ \operatorname{cs} $ function: 
	\begin{align}
		k'(z)=\frac{1}{\sqrt{m}}\left( -\operatorname{cs}^2(2\mathrm{i}z)-\frac{E}{K} \right)=\frac{1}{2\sqrt{m}}\left( \operatorname{dn}^2(2\mathrm{i}z+\mathrm{i}K')+\operatorname{dn}^2(2\mathrm{i}z-\mathrm{i}K')-\frac{2E}{K} \right). \label{eq:kprimez}
	\end{align}
	Since $ Z'=\operatorname{dn}^2-\frac{E}{K} $, integrating this expression soon gives the Jacobi zeta function. Furthermore, the constant of integration can be fixed using some specific value at any point, e.g., $ k_1(\frac{K'}{2})=-\frac{\pi}{4K\sqrt{m}} $. Thus we obtain Eq.~(\ref{eq:kz}).\\
	\indent The third and fourth components of the eigenfunction, $ g_3=\lambda^{-1}\left( -\operatorname{sn}(\frac{x}{\sqrt{m}})g_1-\operatorname{cn}(\frac{x}{\sqrt{m}})g_2 \right) $ and $ g_4=\lambda^{-1}\left( \operatorname{cn}(\frac{x}{\sqrt{m}})g_1-\operatorname{sn}(\frac{x}{\sqrt{m}})g_2 \right) $, can be simplified using another expression of Eq.~(\ref{eq:lambdaz}), $ \lambda(z)=-\mathrm{i} \frac{\Theta_1(\mathrm{i}z)\Theta_2(\mathrm{i}z)}{\Theta_3(\mathrm{i}z)\Theta_4(\mathrm{i}z)} $, and the addition formula of theta functions, in particular, Ref.~\cite{Kharchev201519}, Eqs.~(3.5b) and (3.8). The final expression is then given by Eq.~(\ref{eq:f0xz}).
\section{Completeness relation}\label{app:completeness}
	\indent In this appendix, we derive the completeness relation (\ref{eq:completeness}) for eigenfunctions of the soliton lattice potential $ \phi_0(x) $ of Eq.~(\ref{eq:phi0}).\footnote{
	In order to eliminate confusion, we should explain the difference of the usage of the terminology ``eigenvalues an eigenfunctions'' between classical integrable systems and other fields. In theoretical physics, particularly in quantum mechanics, when one finds a solution of an eigenequation of a given differential operator $ \hat{L}f=\lambda f $ , the operand function $ f $ is called an eigenfunction only when it is a bounded function and only in this case $ \lambda $ is included to the set of eigenvalues. The eigenfunctions possessing everywhere-bounded plane-wave type behavior are called a scattering state and constitute the set of continuous eigenvalues, i.e., a band. The eigenfunction with localized profile and finite norm is called a bound state and make a discrete eigenvalue. The completeness relation, which is in bra-ket notation of quantum mechanics often expressed as $ 1=\sum_n \ket{n}\bra{n} $, is constructed by gathering all these scattering and bound eigenstates.\\
	\indent In classical integrable systems, the exponentially divergent solution of an eigenequation, which is usually not counted as an eigenfunction, plays an important role in generating multi-soliton solutions by various methods. An elementary example is the divergent solution of the Schr\"odinger operator $ -f_{xx}=-\kappa^2 f $ with eigenvalue $ \lambda=-\kappa^2<0 $ and $ f=e^{\pm\kappa x} $, which is indeed used to construct the multi-soliton solution of the KdV equation. Thus, identification of all solutions of the eigenequation for any complex $ \lambda $ is important, and therefore, in the context where no confusion occurs, these divergent solutions $ f $ and corresponding values $ \lambda $ are also sometimes called eigenfunctions and eigenvalues, without distinction. This manuscript also adopts this loose use of terminology in several sections, e.g., in Sec.~\ref{sec:solitonlattice}. To preserve the logical unambiguity, the ``genuine'' eigenfunctions with no divergent behaviors are explicitly referred to as scattering and bound eigenstates. The completeness relation includes only these ``genuine'' eigenfunctions. The solution of eigenequation for arbitrary complex eigenvalue is also often called the Baker-Akhiezer function, whose original meaning is the single-valued function defined on a Riemann surface and possessing finitely many essential singular points but meromorphic excepting these points.
	\indent  
	}\\
	\indent The first important point is that  $ \hat{L} $ is not self-adjoint, i.e., $ \hat{L}^\dagger\ne\hat{L} $ and instead satisfies $ \hat{L}^\dagger=\sigma\hat{L}\sigma $ with $ \sigma=I_2\otimes \sigma_3 $ as stated in Sec.~\ref{sec:laxsymsigmaortho}. For such operator, the eigenfunction satisfies the orthogonal relation with respect to the indefinite inner product $ (f,g)_\sigma = \int dx f^\dagger \sigma g $. The normalization of eigenfunction is also made based on $ (f,f)_\sigma $ and all eigenfunctions are classified into the ones possessing positive, negative, and zero norm. The completeness relation for the case of finite-dimensional linear algebra is shown in section 3 of Ref.~\cite{Takahashi2015101}, and we now need to consider its analog in continuous space for a differential operator. (A specific example for a differential operator in continuous space is found in Refs.~\cite{PhysRevB.91.184501,PhysRevA.96.023626}, though not fully general.) In mathematical physics, a space equipped with indefinite inner product is called the Krein space. Eigenfunctions with nonzero norm can be almost analogously treated to the case of self-adjoint operators whose eigenvalues are real and eigenfunctions are normalized by positive-definite norm. On the other hand, we must carefully consider eigenfunctions with zero norm. In particular, the eigenfunction with complex eigenvalue always has zero norm due to Eq.~(\ref{eq:constJ}). In this case, in order to prepare a $\sigma$-orthogonal basis, we should construct positive- and negative-norm functions by linear combination of a pair of zero-norm eigenfunctions with mutually complex conjugate eigenvalues. \\
	\indent The second important point in writing down the completeness relation supposed to be applied in the ISM is that the integrand must be expressed by meromorphic function with respect to spectral variable $ \lambda $ or its parametrization variable $ z $, because, in the derivation of the GLM equation in \ref{app:glm}, we want to use the residue theorem. Therefore, we must eliminate $ z^* $ from the integrand using the complex conjugation relations described in Eqs.~(\ref{eq:lambdasymm})-(\ref{eq:f0ccrelation}). 
	The key relation is as follows.  A little calculation using the addition formula of theta functions shows
	\begin{align}
		f_0(x,z^*)^\dagger \sigma f_0(x,z)=\sum_{i=1}^4 (-1)^{i-1}g_i(x,z^*)^*g_i(x,z)=\frac{-1}{\sqrt{m}} \left( \operatorname{dn}^2(\tfrac{x}{\sqrt{m}})+\operatorname{cs}^2(2\mathrm{i}z) \right). \label{eq:apcmp1}
	\end{align}
	This quantity is meromorphic, i.e., only including $ z $ due to Eq.~(\ref{eq:f0ccrelation}). The spatial average of this quantity is, using $ \overline{\operatorname{dn}^2}=\frac{E}{K} $ and recalling Eq.~(\ref{eq:kprimez}),
	\begin{align}
		\overline{f_0(x,z^*)^\dagger \sigma f_0(x,z)}=\frac{1}{\sqrt{m}}\left(-\frac{E}{K}-\operatorname{cs}^2(2\mathrm{i}z)\right)=k'(z), \label{eq:normalizationfactor}
	\end{align}
	which is used to normalize scattering eigenstates and rewrite the $ k $-integral to $ z $-integral. As shown below, the spatial average of the norm density $ \overline{f_0(x,z)^\dagger \sigma f_0(x,z)} $ for scattering states all reduces to Eq.~(\ref{eq:normalizationfactor}).\\
	\indent Thirdly, if we have a $ \sigma $-orthogonal basis of the Bloch-type functions parametrized by crystal momentum $ k $ and band index $ \alpha $, and all of them have non-vanishing norm and are normalized so that the spatial average becomes $ \overline{\phi_\pm(k,\alpha)^\dagger \sigma \phi_\pm(k,\alpha)}=\pm 1 $, where $ \pm $ represents the sign of norm, the completeness relation is given by $ \sum_\alpha\int\frac{dk}{2\pi}(\phi_+(k,\alpha)\phi_+(k,\alpha)^\dagger-\phi_-(k,\alpha)\phi_-(k,\alpha)^\dagger)\sigma=\text{Id} $. This fact can be proved by considering the finite-length system where the scattering states can be explicitly normalizable and their spectrum is discretized, and then taking the infinite-length limit. Here we omit the detail of this argument, since it is tedious but straightforward. For readers' reference, we note that a similar discussion in another physical problem for discretization of scattering states of a self-adjoint operator and its infinite limit can be found in Ref.~\cite{takahashinittaJLTP}, section 3.\\
	\indent Keeping in mind the above-mentioned things, let us now write down the completeness relation. Since the potentials (\ref{eq:phi0cossin}) are periodic, there are only scattering states by the Bloch theorem and no bound state exist. Hence, the completeness relation is written only by scattering states. As stated in Sec.~\ref{sec:solitonlattice}, up to periodicity, there are four distinct lines in $z$-plane where crystal momentum becomes real $ k(z)\in\mathbb{R} $, representing the scattering states. Here we discuss the scattering states on each line in detail. 
	\begin{enumerate}[(i)]
	\item $ z\in\mathbb{R} $. \\
	We can check $ \lambda(z)\in\mathbb{R} $ and $ f_0(x,z) $ has positive norm, so $ f_0(x,z^*)=f_0(x,z) $ and hence $ \overline{f_0(x,z^*)^\dagger \sigma f_0(x,z)}=\overline{f_0(x,z)^\dagger \sigma f_0(x,z)}=k'(z)>0 $. Therefore, the contribution of these states to completeness relation is
	 $ \int \frac{dk}{2\pi} \frac{f_0(x,z)f_0(y,z)^\dagger \sigma}{k'(z)}=\int_{-K'}^{K'}\frac{dz}{2\pi} f_0(x,z)f_0(y,z^*)^\dagger \sigma $.
	\item $ z\in\mathbb{R}-\mathrm{i}K $. \\
	We can check $ \lambda(z)\in\mathbb{R} $ and $ f_0(x,z) $ has negative norm, and $ f_0(x,z^*)=-f_0(x,z) $ using periodicity (\ref{eq:f0periodicity}). Thus,  $ \overline{f_0(x,z)^\dagger \sigma f_0(x,z)}=-\overline{f_0(x,z^*)^\dagger \sigma f_0(x,z)}=-k'(z)<0 $. Therefore, the contribution to these states to completeness relation is 
	$ \int dk \frac{f_0(x,z)f_0(y,z)^\dagger \sigma}{-k'(z)}=\int_{-K'-\mathrm{i}K}^{K'+\mathrm{i}K}\frac{dz}{2\pi} f_0(x,z)f_0(y,z^*)^\dagger \sigma. $ 
	\item $z\in\mathbb{R}\pm\frac{\mathrm{i}K}{2}$. \\
	We can check $ \lambda(z)\in\mathrm{i}\mathbb{R} $ and hence $ f_0(x,z) $ has zero norm. Therefore we must make nonzero-norm functions by linear combination of a pair of eigenfunctions with mutually complex conjugate eigenvalues. Since $ \lambda(z^*)=\lambda(z)^* $ and $ k(z^*)=k(z)^* $, this pair share the same wavenumber.  Now let us assume that $ f_1 $ and  $ f_2 $ are zero-norm eigenfunctions of the Bloch type with complex eigenvalues $ \lambda $ and $ \lambda^* $ and share the same and real-valued wavenumber $ k $. By definition the spatial average of norm density is zero: $ \overline{f_i^\dagger \sigma f_i}=0,\ i=1,2 $, but $ \overline{f_1^\dagger\sigma f_2}\ne 0 $ unless they belong to nontrivial Jordan blocks. Then, if an overall prefactors of $ f_1 $ and $ f_2 $ are adjusted to satisfy the relation $ \overline{f_1^\dagger\sigma f_2}=\overline{f_2^\dagger\sigma f_1}>0 $, then $ f_\pm=\frac{f_1\pm f_2}{\sqrt{2}}$ are positive- and negative-norm functions and $ \sigma $-orthogonal to each other.  Thus these two can be used as a $\sigma$-orthogonal basis, and the contribution of these states to completeness relation is written as $ \int\frac{dk}{2\pi}\frac{f_+f_+^\dagger\sigma-f_-f_-^\dagger\sigma}{\overline{f_2^\dagger\sigma f_1}}=\int\frac{dk}{2\pi}\frac{(f_1f_2^\dagger+f_2f_1^\dagger)\sigma}{\overline{f_2^\dagger\sigma f_1}}. $ Now let us take $ f_1=f_0(x,z),\ z\in\mathbb{R}+\frac{\mathrm{i}K}{2} $ and $ f_2=-f_0(x,z^*),\ z^*\in\mathbb{R}-\frac{\mathrm{i}K}{2} $. By Eq.~(\ref{eq:normalizationfactor}), $ \overline{f_2^\dagger\sigma f_1}=-k'(z)>0 $ is real and positive. (Note: $ k(z) $ is a decreasing function on $ z\in\mathbb{R}\pm\frac{\mathrm{i}K}{2} $.) Therefore, the contribution of these states to completeness relation is given by $ \int\frac{dk}{2\pi}\frac{(-f_0(x,z)f_0(y,z^*)^\dagger-f_0(x,z^*)f_0(y,z)^\dagger)\sigma}{-k'(z)}=-\int_{-K'+\frac{\mathrm{i}K}{2}}^{K'+\frac{\mathrm{i}K}{2}}\frac{dz}{2\pi}(f_0(x,z)f_0(y,z^*)^\dagger+f_0(x,z^*)f_0(y,z)^\dagger)\sigma $. Changing the dummy variable $ z\to z^* $, the latter term can be rewritten as $ -\int_{-K'-\frac{\mathrm{i}K}{2}}^{K'-\frac{\mathrm{i}K}{2}}\frac{dz}{2\pi}f_0(x,z)f_0(y,z^*)^\dagger\sigma $. 
	\end{enumerate}
	Summarizing all contributions from (i)-(iii), the completeness relation is given by
	\begin{align}
		\delta(x-y)I_4&=\int_{-K'}^{K'}\frac{dz}{2\pi} f_0(x,z)f_0(y,z^*)^\dagger \sigma+\int_{-K'-\mathrm{i}K}^{K'+\mathrm{i}K}\frac{dz}{2\pi} f_0(x,z)f_0(y,z^*)^\dagger \sigma\nonumber \\
		&\qquad-\int_{-K'+\frac{\mathrm{i}K}{2}}^{K'+\frac{\mathrm{i}K}{2}}\frac{dz}{2\pi}f_0(x,z)f_0(y,z^*)^\dagger\sigma-\int_{-K'-\frac{\mathrm{i}K}{2}}^{K'-\frac{\mathrm{i}K}{2}}\frac{dz}{2\pi}f_0(x,z)f_0(y,z^*)^\dagger\sigma.
	\end{align}
	Adding vertical paths which cancel out by periodicity, we obtain the desired form
	\begin{align}
		\delta(x-y)I_4 = \oint_{C_1+C_2}\frac{dz}{2\pi} f(x,z)f(y,z^*)^\dagger\sigma,
	\end{align}
	where the definitions of the contours $ C_1 $ and $ C_2 $ are already described in the sentences after Eq.~(\ref{eq:completeness}).
\section{Derivation of the GLM equation}\label{app:glm}
	In this appendix we drive the GLM equation (\ref{eq:glm}). By the relation between right and left Jost solutions (\ref{eq:scatteringmatrix}) and the integral representation (\ref{eq:kernelrep}),
	\begin{align}
		f_+(x,z)\frac{1}{a(z)} -f_0(x,z)=\int_{-\infty}^x \mathrm{d}y \Gamma(x,y)f_0(y,z)+\left( f_0(x,-z-\mathrm{i}K)+\int_{-\infty}^x\mathrm{d}y \Gamma(x,y)f_0(y,-z-\mathrm{i}K) \right)\frac{b(z)}{a(z)}.
	\end{align}
	Multiplying $ \frac{f(w,z^*)^\dagger\sigma}{2\pi},\ w \le x $ to both sides of the above equation from right, we integrate the expression by $ z $ along the contour $ C_1+C_2 $.
	First, let us evaluate the right-hand side. Defining
	\begin{align}
		\Omega_{\text{sc}}(x,w):=\int_{C_1+C_2} \frac{\mathrm{d}z}{2\pi} f_0(x,-z-\mathrm{i}K)\frac{b(z)}{a(z)}f_0(w,z^*)^\dagger\sigma,
	\end{align}
	and using the completeness relation (\ref{eq:completeness}), we find
	\begin{align}
		\text{R.H.S.}= \Gamma(x,w)+\Omega_{\text{sc}}(x,w)+\int_{-\infty}^x\mathrm{d}y \Gamma(x,y) \Omega_{\text{sc}}(y,w).
	\end{align}
	In the contour integral of $ \Omega_{\text{sc}} $, actually the integrals for scattering states only contribute, and hence it can be rewritten as Eq.~(\ref{eq:omegasc}).\\
	\indent Next, let us evaluate the left-hand side. In the SG equation, generally $ a(z) $ can have higher-order zeros \cite{doi:10.1143/JPSJ.53.2908}, but here we only consider the case where all zeros are of the first order. Let zeros of $ a(z) $ be $ z_1,\dots,z_n $. As already mentioned before, by Eq.~(\ref{eq:scatinvo2}), the zeros of $ a(z) $ simultaneously emerge at $ z,-z^*, z-\mathrm{i}K, -\mathrm{z}^*-\mathrm{i}K $. Therefore, not all $ z_j $'s can be freely chosen. Here, however, we formally assign independent labels to all zeros. Then,  the left-hand side can be calculated using the residue theorem:
	\begin{align}
		\text{L.H.S.}&=\sum_{j=1}^n \frac{\mathrm{i}f_+(x,z_j)f_0(w,z_j^*)^\dagger}{\dot{a}(z_j)} \nonumber \\
		&=\sum_{j=1}^n \frac{\mathrm{i}f_-(x,-z_j-\mathrm{i}K)b(z_j)f_0(w,z_j^*)^\dagger}{\dot{a}(z_j)},
	\end{align}
	where the relation $ f_+(x,z_j)=b(z_j)f_-(x,-z_j-\mathrm{i}K) $ arising from the condition $ a(z_j)=0 $ has been used.  
	If we define
	\begin{align}
		\Omega_{\text{bd}}(x,w):=\sum_{j=1}^n f_0(x,-z_j-\mathrm{i}K)c_j^2f_0(w,z_j^*)^\dagger\sigma,\quad c_j^2:=\frac{b(z_j)}{\mathrm{i}\dot{a}(z_j)}, \label{eq:omegabd0}
	\end{align}
	then
	\begin{align}
		\text{L.H.S.}=-\Omega_{\text{bd}}(x,w)-\int_{-\infty}^x\mathrm{d}y\Gamma(x,y)\Omega_{\text{bd}}(y,w).
	\end{align}
	To summarize, writing $ \Omega=\Omega_{\text{sc}}+\Omega_{\text{bd}} $, we obtain the final result
	\begin{align}
		\Gamma(x,w)+\Omega(x,w)+\int_{-\infty}^x\mathrm{d}y\Gamma(x,y)\Omega(y,w)=0,\quad w\le x.
	\end{align}
	If we simplify Eq.~(\ref{eq:omegabd0}) using Eqs.~(\ref{eq:f0periodicity}) and (\ref{eq:f0ccrelation}), it is rewritten as Eq.~(\ref{eq:omegabd}) with (\ref{eq:omegabd2}).

\section{Calculation of determinant}\label{app:dettheta}
	Here we calculate the asymptotic form of the multi-soliton solution for $ x\to+\infty $. In particular, we provide the shift of the lattice $ x_0 $ in Eq.~(\ref{eq:x0asymp}).
	First, we write $ \mathcal{P}=S_1\tilde{\mathcal{P}}S_2 $ and $ \mathcal{Q}=S_3\tilde{\mathcal{Q}}S_4 $, where
	\begin{alignat}{2}
		S_1&=\operatorname{diag}\left( \frac{\Theta_4(\mathrm{i}z_1)}{\Theta_1(\mathrm{i}z_1)},\dots,\frac{\Theta_4(\mathrm{i}z_n)}{\Theta_1(\mathrm{i}z_n)} \right),\quad &S_2&=\operatorname{diag}\left( \frac{\Theta_2(\mathrm{i}z_1)}{\Theta_3(\mathrm{i}z_1)},\dots,\frac{\Theta_2(\mathrm{i}z_n)}{\Theta_3(\mathrm{i}z_n)} \right), \\
		S_3&=\operatorname{diag}\left( \frac{\Theta_4(\mathrm{i}z_1)}{\Theta_2(\mathrm{i}z_1)},\dots,\frac{\Theta_4(\mathrm{i}z_n)}{\Theta_2(\mathrm{i}z_n)} \right),\quad &S_4&=\operatorname{diag}\left( \frac{\Theta_1(\mathrm{i}z_1)}{\Theta_3(\mathrm{i}z_1)},\dots,\frac{\Theta_1(\mathrm{i}z_n)}{\Theta_3(\mathrm{i}z_n)} \right),
	\end{alignat}
	and
	\begin{align}
		\tilde{\mathcal{P}}_{ij}&=\frac{-\Theta_4\Theta_2(\frac{x}{\sqrt{m}}+\mathrm{i}(z_i+z_j))}{\Theta_2(\mathrm{i}(z_i+z_j))\Theta_1(\frac{x}{\sqrt{m}})},\\
		\tilde{\mathcal{Q}}_{ij}&=\frac{\Theta_4\Theta_1(\frac{x}{\sqrt{m}}+\mathrm{i}(z_i+z_j))}{\Theta_2(\mathrm{i}(z_i+z_j))\Theta_2(\frac{x}{\sqrt{m}})}.
	\end{align}
	Then, using the formula in Ref.~\cite{PhysRevE.93.062224}, appendix D, we obtain
	\begin{align}
		\frac{\det \tilde{\mathcal{P}}}{\det\mathcal{G}}&=(-1)^n\frac{\Theta_4(\frac{x}{\sqrt{m}})\Theta_1(\frac{x}{\sqrt{m}}+nK+2\mathrm{i}\sum_{j=1}^n z_j)}{\Theta_1(\frac{x}{\sqrt{m}})\Theta_4(\frac{x}{\sqrt{m}}+nK+2\mathrm{i}\sum_{j=1}^n z_j)}, \\
		\frac{\det \tilde{\mathcal{Q}}}{\det\mathcal{G}}&=(-1)^n\frac{\Theta_4(\frac{x}{\sqrt{m}})\Theta_2(\frac{x}{\sqrt{m}}+nK+2\mathrm{i}\sum_{j=1}^n z_j)}{\Theta_2(\frac{x}{\sqrt{m}})\Theta_4(\frac{x}{\sqrt{m}}+nK+2\mathrm{i}\sum_{j=1}^n z_j)}.
	\end{align}
	This relation itself is generally valid even when all $ z_j $'s are independent and $ n $ is not even. \\
	As discussed in Sec.~\ref{sec:realcondition}, for the real and bounded solution of the SG equation,  $ n $ is even and  $ z_j $ and $ z_j-\mathrm{i}K $ simultaneously appear. In this case, we label the zeros, e.g., as $ z_{n/2+j}=z_j-\mathrm{i}K, $  $  j=\frac{n}{2}+1,\dots, n $. Then, $ \det S_1=\prod_{j=1}^{n/2}\frac{\Theta_4(\mathrm{i}z_j)\Theta_3(\mathrm{i}z_j)}{\Theta_1(\mathrm{i}z_j)\Theta_2(\mathrm{i}z_j)},\quad \det S_2=(-1)^{n/2}\det S_1^{-1},\quad \det S_4=\prod_{j=1}^{n/2}\frac{\Theta_1(\mathrm{i}z_j)\Theta_2(\mathrm{i}z_j)}{\Theta_3(\mathrm{i}z_j)\Theta_4(\mathrm{i}z_j)},\quad \det S_3=(-1)^{n/2}\det S_4^{-1} $, and hence
	\begin{align}
		\frac{\det \mathcal{P}}{\det\mathcal{G}}&=\frac{\Theta_4(\frac{x}{\sqrt{m}})\Theta_1(\frac{x}{\sqrt{m}}+2\mathrm{i}\sum_{j=1}^n z_j)}{\Theta_1(\frac{x}{\sqrt{m}})\Theta_4(\frac{x}{\sqrt{m}}+2\mathrm{i}\sum_{j=1}^n z_j)}=\frac{\cos\frac{\phi_0(x-x_0)}{2}}{\cos\frac{\phi_0(x)}{2}}, \\
		\frac{\det \mathcal{Q}}{\det\mathcal{G}}&=\frac{\Theta_4(\frac{x}{\sqrt{m}})\Theta_2(\frac{x}{\sqrt{m}}+2\mathrm{i}\sum_{j=1}^n z_j)}{\Theta_2(\frac{x}{\sqrt{m}})\Theta_4(\frac{x}{\sqrt{m}}+2\mathrm{i}\sum_{j=1}^n z_j)}=\frac{\sin\frac{\phi_0(x-x_0)}{2}}{\sin\frac{\phi_0(x)}{2}}.
	\end{align}

\bibliographystyle{ptephy_authornameperiod_nomonth}

{\small
%\bibliography{SG5}

}

\end{document}